\documentclass[10pt,twocolumn,letterpaper]{article}

\usepackage{cvpr}
\usepackage{times}
\usepackage{epsfig}
\usepackage{graphicx}
\usepackage{amsmath}
\usepackage{amssymb}
\usepackage{tabularx}
\usepackage{adjustbox}

\usepackage[breaklinks=true,bookmarks=false]{hyperref}

\cvprfinalcopy 

\def\cvprPaperID{****} 

\begin{document}

\title{Homomorphic Encryptions for Privacy Preserving Vision}

\author{Preey Shah\\
{\tt\small preey@stanford.edu}
\and
Rohan Virani\\
{\tt\small rohan99@stanford.edu}
\and
Sanjari Srivastava\\
{\tt\small sanjari4@stanford.edu}
}

\maketitle

\begin{abstract}
Legal requirements might prevent organizations from sharing sensitive data like medical or financial details of consumers which prevents them from leveraging cloud based ML-as-a-service solutions provided by third party providers, which are quickly gaining popularity these days. In this project, we aim to perform inference tasks in Computer Vision in a privacy-preserving manner, i.e, by only looking at encrypted data. Recent advances in fully homomorphic encryption make this possible. A fully homomorphic encryption allows an arbitrary sequence of additive and multiplicative operations to be performed on encrypted data directly. Applying homomorphic encryptions to CNNs requires modifying the conventional CNN layers, so that they adhere to the encryption scheme.

Our aim was to explore the best methods to create CNNs which can classify encrypted images directly. We used Microsoft SEAL \cite{sealcrypto} for performing homomorphic encryption. The performance of these "encryption based CNNs" should be comparable with baseline accuracies of the same CNNs trained on unencrypted data, and the aim was to achieve as low of a hit on inference-time performance as possible. 

We successfully obtained minimal drop in classification accuracy for various datasets. We used MNIST as our baseline, which is popularly used in related research work \cite{https://doi.org/10.48550/arxiv.1412.6181}\cite{tenseal2021} and then explored more complex datasets like Kuzushiji MNIST, Fashion-MNIST and CIFAR-10 as a part of our contribution. 

Additionally, we also added support for more complex operations on top of TenSEAL \cite{tenseal2021}, like processing colored images (multi-channel input), applying multiple convolutional layers and performing average pooling. 
\end{abstract}

\section{Introduction}
Deep Learning is heavily used for common computer vision tasks like optical character recognition, segmentation and facial recognition. However, recently computer vision is also finding increasing uses in fields like healthcare, for tasks like tumor detection \cite{brain_tumor}, medical imaging \cite{Wang_2020_CVPR}, detecting skin cancer \cite{DBLP:journals/corr/LiEKNKT16}. Applying machine learning, especially to problems involving medical/financial or other type of sensitive data, requires careful maintenance of data privacy and security. At the same time, the \textit{Prediction-as-a-Service} paradigm is also gaining traction, in which a prediction service manages the cloud infrastructure needed to run a model at scale, and makes it available for online and batch prediction requests. Since legal requirements to preserve data privacy may prevent healthcare companies from using cloud-based machine learning solutions, there is a need for neural networks which can be applied directly on encrypted data; without decrypting it on the cloud.

We studied various works which use neural nets with homorphic encryption to achieve privacy-preserving vision, like CryptoNets \cite{https://doi.org/10.48550/arxiv.1412.6181}, PySyft\cite{pysyft} , TenSEAL\cite{tenseal2021}. We use TenSEAL for our experiments, because it is an easy to use python wrapper on top of Microsoft SEAL's C++ implementation. We demonstrate homomorphic operations on encrypted data for datasets like MNIST, Kuzushiji-MNIST, Fashion-MNIST and on the more complex CIFAR-10. We also note how non-linearity in deep neural networks should be modified to allow it to work with homomorphic encryptions. For a baseline comparison, we use the same CNNs on the prediction task on unencrypted data. 

We demonstrate how to perform classification tasks on encrypted data without taking a significant hit on the accuracy and test-time inference speeds. Tuning the encryption scheme's context parameters plays a major role in ensuring this and we empirically analyzed this time-accuracy tradeoff.

We implemented additional useful operations like stacked convolutional layers, multi-channel inputs and Average pooling on top of TenSEAL to enable creating more complex CNN models as a part of our contribution.


\section{Problem Statement}
Our main problem statement is to run deep learning models on encrypted data in computer vision tasks, while trying to ensure minimal loss of accuracy, and reducing inference time. In particular, we wanted to simulate running feature identification on a privacy sensitive problem like recognizing facial features/medical data \cite{liu2015faceattributes} by using complex Computer Vision datasets like CIFAR-10.\\
We note that operations in homomorphic encryption take a significant amount of time to run, and our principal efforts were directed towards approximating deep learning models that do not lose accuracy and accelerate inference decisions. We also explored this time-accuracy tradeoff in detail. 

\section{Related Work}
The Cryptonets paper by Microsoft \cite{https://doi.org/10.48550/arxiv.1412.6181} opened up many possibilities in this field of privacy preserving ML. Later works include progress in federated learning \cite{fl}, that is mostly focused on privacy preservation through combining models that run locally. This progress was made possible by advances in homomorphic encryption \cite{gentry}. \\
Later work in using encrypted data includes \cite{he} that combines ideas of running models locally and using homomorphic encryption. Significant efforts have been made to improve the running times, though this problem does not seem to have applied to other domains. There have been some recent attempts in trying to make this efficient \cite{rw}

Once similar accuracies are achieved on unencrypted and encrypted data, more improvements can then be made to ensure better inference speed at test time by using GPUs and faster homomorphic encryption methods. Related works like \cite{fast_HME}, \cite{fast_hme_f1} explore faster ways to perform Fully Homorphic Encryption which can help achieve this. 

\subsection{Homomorphic Encryption}


Data encryption is a way of translating plaintext into ciphertext in order to maintain security and privacy of data. Homomorphic encryption (HE) (Rivest et al\cite{Rivest1978}) adds to that the ability to perform mathematical operations on the data while it is still encrypted. This means that under HE, the result of performing an operation on two ciphertexts would be the same as performing the operation on the corresponding plaintexts and then encrypting the result.

HE preserves homomorphism over additive and multiplicative operations. The first such encryption scheme was introduced by Gentry et al.\cite{gentry}, and was soon followed by many advances in this field (e.g. Naehrig et al. \cite{naehrig}; Gentry et al. (2013) \cite{crypto-2013-24633} López-Alt et al. (2012) \cite{LpezAlt2012OntheflyMC}). A fully homomorphic encryption should allow for an arbitrary number of addition and multiplication operations to be performed on the encrypted data. 

If $\Phi$ repesents an encryption scheme which converts a plaintext to a cipher, and $\oplus$ and $\otimes$ are the addition and multiplication operations defined on a commutative ring (like the set of integers $\mathbb{Z}$ or $\mathbb{Z}\%m$), then HE requires that:

$$\Phi\left(z_{1}+z_{2}\right)=\Phi\left(z_{1}\right) \oplus \Phi\left(z_{2}\right)$$ and $$\Phi\left(z_{1} \cdot z_{2}\right)=\Phi\left(z_{1}\right) \otimes \Phi\left(z_{2}\right)$$

\subsection{Modifying Neural Networks}
Following common operations are typically present in a convlutional neural network:
\begin{enumerate}
  \item Weighted-Sum (convolution layer): Multiply the vector of values (kernel) at the layer beneath it by a vector of weights and sum the results. The filter or kernel is "convolved" across the input as it reduces all the pixels in its receptive field into a single value. This function is essentially a dot product of the weight vector and the vector of values of the feeding layer.
  \item Max Pooling: Compute the maximal value of some of the components of the feeding layer.
  \item Sigmoid: Take the value of one of the nodes in the feeding layer and evaluate the function $z \mapsto 1 /(1+\exp (-z))$.
  \item Rectified Linear Units (ReLU): Take the value of one of the nodes in the feeding layer and compute the function $z \mapsto$ $\max (0, z)$.
\end{enumerate}

Due to the constraint that HE is defined for additive and multiplicative operations, non-polynomial operations like the $\max()$ function cannot be supported under an HE scheme. Also, HE schemes work on the domains of commutative rings like the set of integers $\mathbb{Z}$. Therefore, the following modifications become necessary to allow our CNN to work with encrypted data (ideas borrowed from CryptoNets (\cite{naehrig}):
\begin{enumerate}
    \item \textbf{Real numbers} are replaced by fixed precision floating point numbers whose binary notations can be encoded into integers.
    \item \textbf{Non-polynomial activation functions} like ReLU and Sigmoid layers are replaced by low degree polynomials, ie, sqr $(z):=z^{2}$ to introduce non-linearity. We also found that the performance of other polynomials like cubic functions for the non-linear layers also leads to comparable results for simple datasets.
    \item \textbf{Max-pooling} is replaced by a scaled mean-pooling layer.
    \item \textbf{Plain operations} Since the weights and biases (W, b) of the CNN remains known to the ML service provider, we do not need to encrypt them before taking dot products. The naive way to implement such operations is to first encrypt these known "constants" and then perform the addition or multiplication operation (the weights of the network can change during training but are frozen during test-time). However, this feed-forward operation can be simplified as follows:
    
    \begin{itemize}
        \item Let $c=\lfloor q / t\rfloor m+e+h s$ be the encrypted message and $w$ the known constant. Addition can be achieved by multiplying $w$ by $\lfloor q / t\rfloor$ and adding that to $c$, which results in $\lfloor q / t\rfloor(m+w)+e+h s$. This is essentially just encrypting $w$ with no noise and performing normal homomorphic addition.
        \item For multiplication, even the scaling is not needed since $c w=\lfloor q / t\rfloor m w+e^{\prime}+h s^{\prime}$. This is very efficient, especially if $w$ is a sparse polynomial. For example, if $w$ is a scalar (as it would be in the scenario below), then this multiplication is computed in linear time in the degree of $c$, which is $n-1$.
    \end{itemize}
\end{enumerate}

Next we describe how a convolution operation can be performed on encrypted data. Although there are several methods for modifying a convolution operation to be vectorized: we use the image to columns method as explained below. For each part of the input that a $f$ x $f$ kernel would pass over to compute a dot product, we flatten the input into a row vector of length $f^{2}$ to be multiplied by the column vector representing the kernel. Thus convolution is represented as a matrix multiplication - which would need to be reshaped back into a square. This step rearranges the elements of the input matrix, however this trick is difficult to do on ciphertext and thus must be done prior to encrypting the data. 

\includegraphics[scale=0.5]{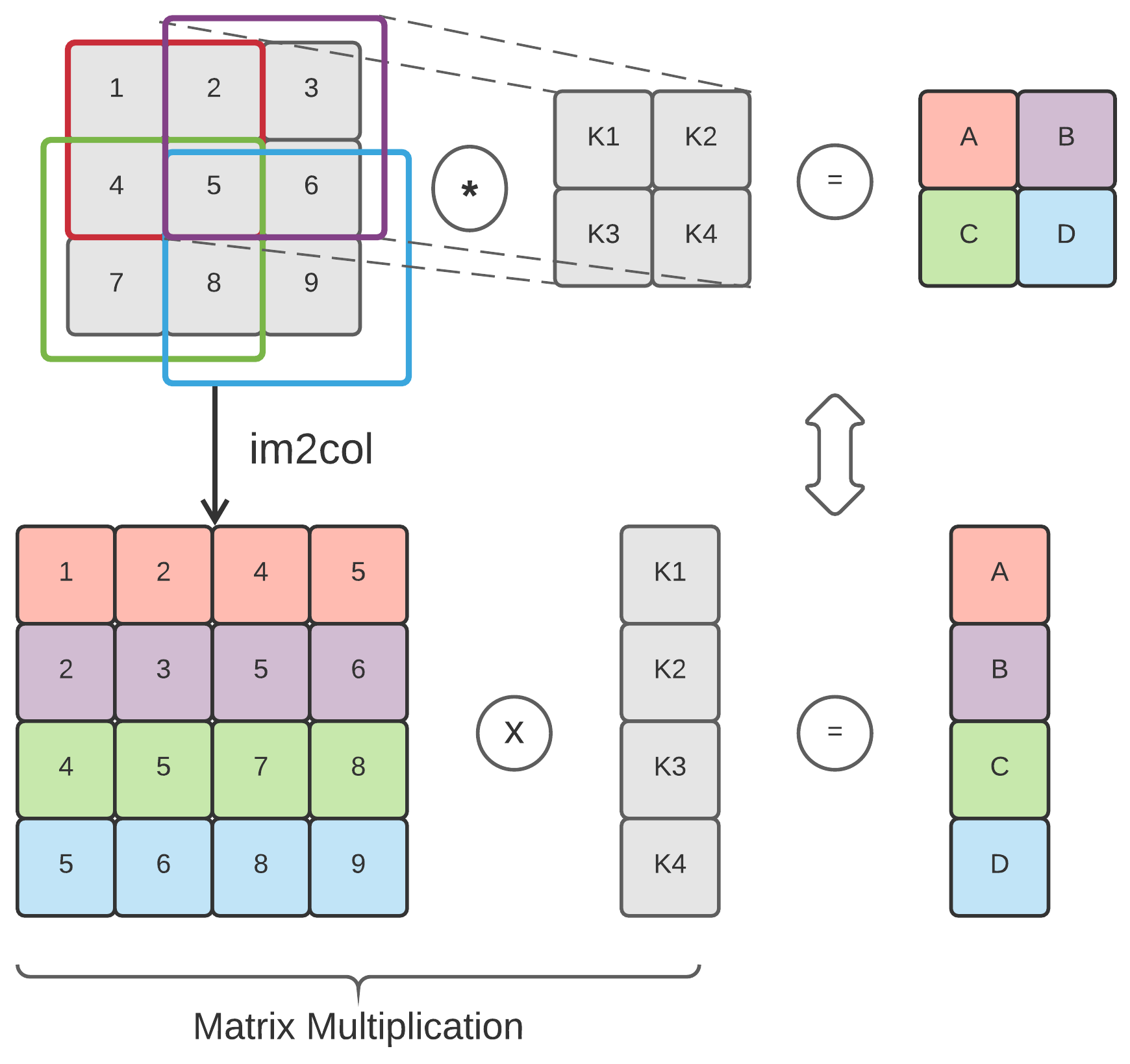}

Let the result of the previous rearrangement be some matrix $A$ with width $f^{2}$. We then flatten the input into a vector by scanning columns from left to right such that if $A_{i}$ is the i'th column of $A$ then the vector is $[A_{1}^{T},...,A_{f^{2}}^{T}]$. Similarly we flatten the convolution kernel by repeating each element of the kernel $n$ times where n is the height of the input matrix $A$. We can now perform an elementwise ciphertext multiplication followed by a series of rotation and sum operations in order to sum elements in the same convolutional window. If multiple kernels are used, then the process above is merely repeated and the resultant vectors are concatenated together (i.e. if 4 kernels were used and the output of one kernel is a vector of size 64, then the actual result will be a vector of size 256). 

\includegraphics[scale=0.5]{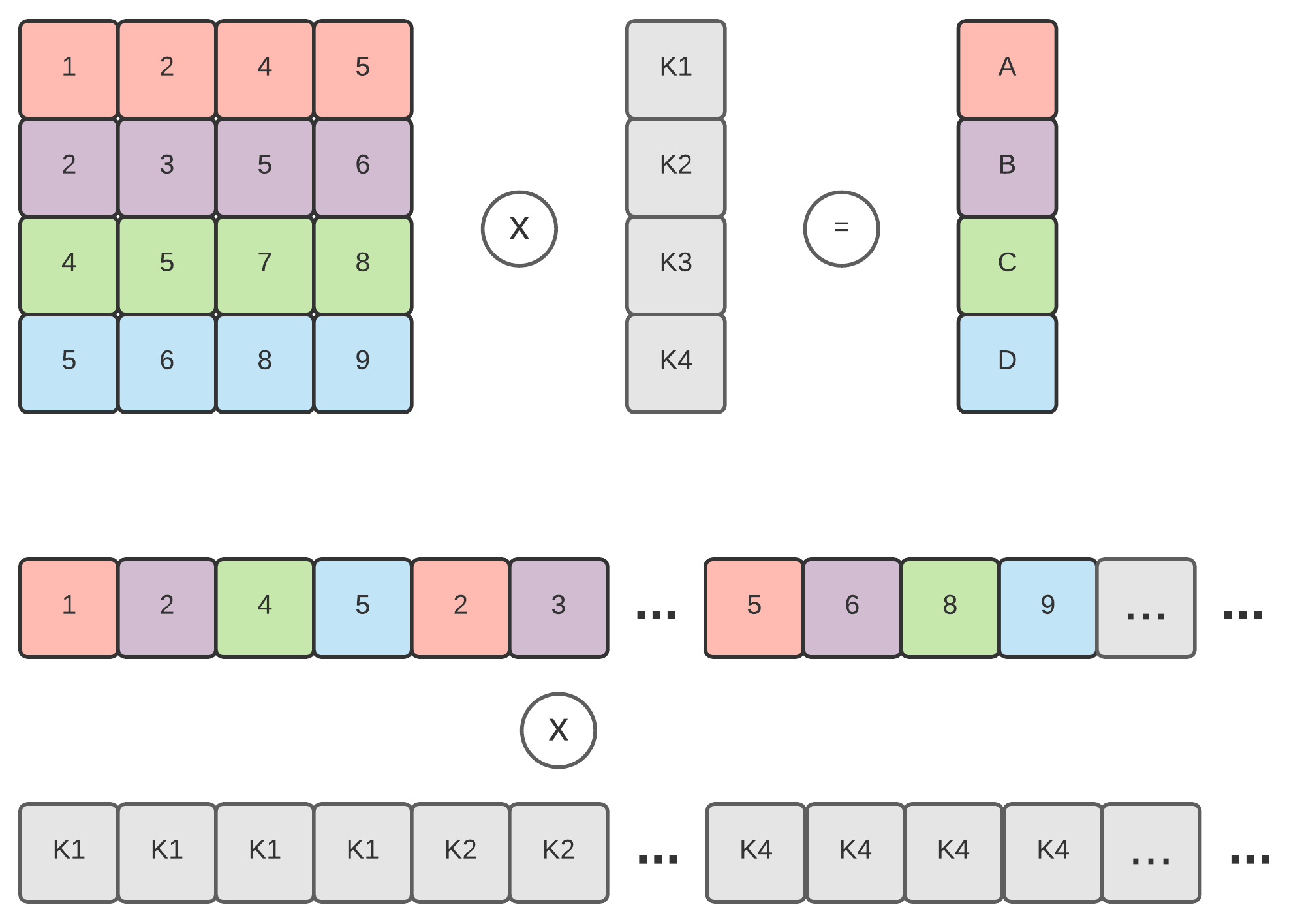}

For the linear operation, we utilize the Halevi and Shoup method for computation in encrypted space. This is a set of ciphertext multiplication operations on a rotated matrix. We iterate over every diagonal in the weight matrix and multiply it by ciphertext rotated n slots to the left. This is better shown by the diagram below. 

\includegraphics[scale=0.16]{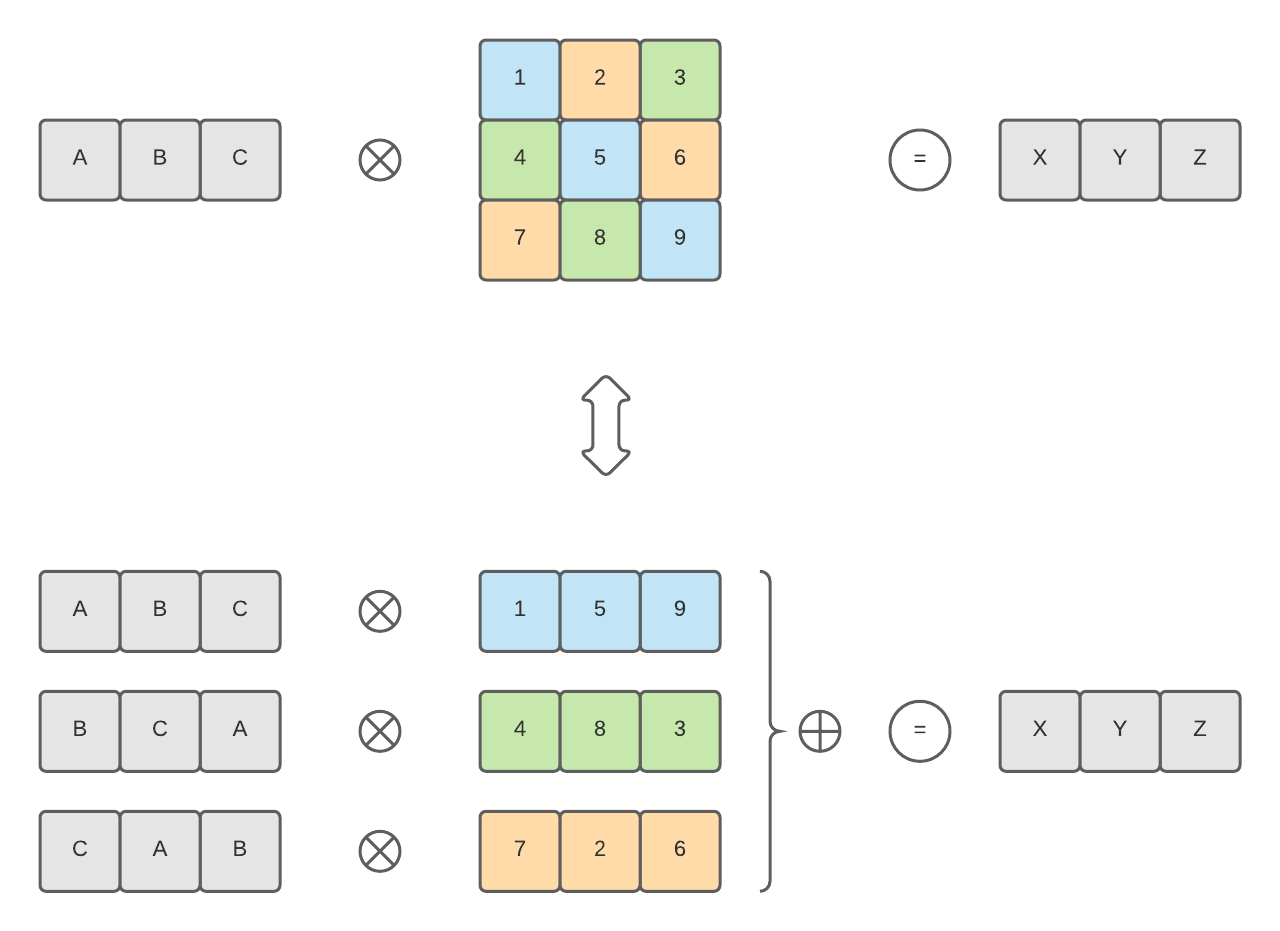}

Finally, the method we use for pooling is using average pooling. This is equivalent to a convolution operation with a static, unoptimized kernel and thus the method for doing this on encrypted data is exactly the same as the image to columns method described earlier. Given that we must rearrange the input matrix for a convolution or pooling operation prior to encryption, this implies that we cannot do more than one convolution on encrypted data. We circumvent this by decrypting and then re-encrypting the data prior to each convolutional layer. This theoretically presents no issues for data privacy since the only information necessary for rearranging and then encrypting data is kernel size and stride, not the existing model parameters. 

\subsection{CKKS Encryption}

For the purposes of this project, we use the Cheon-Kim-Kim-Song homomorphic encryption scheme [CKKS17] encryption scheme. Unlike other schemes such as BFV which use exact arithmetic such that the result is the same as the input after decryption, CKKS uses approximate arithmetic. We will explain at a high level how the CKKS encryption and decryption scheme works. We first consider message defined in the space of $\mathbb{C}^{\frac{N}{2}}$ and encode this into plaintext in the space of $\mathbb{Z}[X]/(X^{N}+1)$. This is then encrypted to using a public key to produce ciphertext in the space of $(\mathbb{Z}[X]/(X^{N}+1))^{2}$. We can then conduct computation on this ciphertext followed by a decryption and decoding operation to retrieve the result in the original vector space.


\section{Datasets}
We decided to use MNIST as a baseline for our experiments since it is simple and is commonly used in related works in the field of encryption based ML.

Additionally, we explored newer datasets
like Kuzushiji MNIST, Fashion-MNIST and CIFAR-10 which are of a similar size but with a much higher complexity than MNIST. 
It is extremely easy to access and preprocess these datasets using the TorchVision library \cite{torchvision}. To pre-process our data, we normalized it and converted CIFAR-10 images to grayscale as needed depending upon the architecture we were testing.

All above-mentioned datasets contain 60k images. The MNIST datasets contain grayscale images (28x28 pixels) while CIFAR-10 contains RGB colored images (3x32x32 pixels). We trained all models on the training split of 50k images which torchvision provides by default.

We note that the bottleneck for encrypted CNN is the performance of the model during inference. The encrypted data becomes almost one to three orders of magnitude larger than the unencrypted data for the MNIST dataset if we use double precision floating point numbers. Also, adding homomorphic encryption at test time makes the process at least an order of magnitude slower.

Therefore, we restricted our test split to 5k samples for each dataset and used the remaining 5k images as a validation set.

\section{Experimental Setup and Methodology}

\subsection{Encryption Library}
Microsoft SEAL is an open-source homomorphic encryption library which provides a set of encryption libraries that allow computations to be performed directly on encrypted data. We also make use of the open source library TenSEAL (\cite{tenseal2021}) by OpenMined which provides a python API on top  of Microsoft SEAL's C++ implementation making it very easy to use with PyTorch based machine learning models. 

The \href{https://github.com/OpenMined/PySyft}{Syft} ecosystem created by OpenMined, can also be used since it provides secure and private Deep Learning in Python and implements other methods of security as well like Federated Learning, Differential Privacy, and Encrypted Computation (like Multi-Party Computation (MPC) along with Homomorphic Encryption (HE)), however we conducted all our experiments using TenSEAL due to its ease of use.

\subsection{Evaluation Metrics}

We compared the different encrypted-CNN models on the following metrics.
\begin{itemize}
    \item Classification accuracy
    \item Inference time per encrypted instance
\end{itemize}

We also varied the models on the following axes during our evaluation.
\begin{itemize}
    \item Training on different computer-vision datasets
    \item Implementing complex layers like stacked convolutional layers, average pooling and colored inputs
    \item Varying encryption scheme related parameters, like the floating point precision to be used for encoding real numbers in the CNNs.
\end{itemize}

\subsection{"Encrypted" CNNs}
We performed encrypted inference on the 3 CNN models which differ from each other in the complexity of the layers used. 

\emph{We emphasize that these are really basic neural networks, because the idea was not to beat state of the art results, but to show that using these models as baselines, we can obtain comparable test accuracies on encrypted images as well.}

Each model is trained on unencrypted training data first. The trained weights are then directly used to initialize a ciphertext-supportive variant of the model which can work on encrypted images. This encryption-based model substitutes each additive/multiplicative operation by the analogous homomorphic operator.

Thus the network weights don't get encrypted but this "encrypted" variant of the model can obtain classification results on fully encrypted images.

We have highlighted the non trivial layers we added in each architecture. All the convolutional layers had a padding of 0. 
\subsubsection{Architecture 1}
The first setup we used was a simple CNN with one convolutional layer and 2 FC layers. The input image was also constrained to a single channel. 
\vspace{1em}\\
\fbox{
\begin{minipage}{20em}
\textbf{Conv2d(in\_ch=1},out\_ch=4,kernel=7,stride=3)\\
\textbf{SquaredLayer()}\\
Linear(\_, 64) \\
\textbf{SquaredLayer()}\\
Linear(64, 10)
\end{minipage}
}

\subsubsection{Architecture 2}

We implemented multiple convolutional layers and average pooling in TenSEAL. [Explained in detail in Section 3.2]

The stacked convolutional layers are implemented by decrypting and reencrypting the output between each layer. We do this because TenSEAL does not provides documentation around dealing with 2-D matrices which makes rearranging the encrypted 1-D vector non trivial after a convolution. However, this could also be potentially be done without the decryption/reencryption phase by using CKKSTensors.

Average pooling is implemented by treating it like any other convolution with a fixed kernel where each value is set to 0.25 for a 2x2 kernel. 

This allowed us to use the second CNN with two convolutional layers, average pooling and 2 FC layers. The input image was constrained to be single-channel. \\
\vspace{1em}\\
\fbox{
\begin{minipage}{20em}
Conv2d(in\_ch=1,out\_ch=1,kernel=4,stride=2)\\
\textbf{AveragePool()}\\
\textbf{Conv2d(in\_ch=1,out\_ch=16,kernel=3,stride=1)}\\
SquaredLayer()\\
Linear(\_, 64) \\
SquaredLayer()\\
Linear(64, 10)
\end{minipage}
}

\subsubsection{Architecture 3}
Finally we implemented support for multiple channel input for colored images. 

The outputs of the convolutions on each channel were computed separately and then combined with the $ts.ckks\_vector.add()$ operator to directly add the ciphertexts corresponding to each channel. 
For our experiments, we kept the remainder of this model simple for the interest of time and only increased the sizes of the FC layers for better baseline accuracies.
\vspace{1em}\\
\fbox{
\begin{minipage}{20em}
\textbf{Conv2d(in\_ch=3},out\_ch=6,kernel=5,stride=3)\\
SquaredLayer()\\
Linear(600, 200) \\
SquaredLayer()\\
Linear(200, 10)
\end{minipage}
}

\subsection{Unencrypted training phase}
We followed basic guidelines for training of the models on unencrypted data. For all models, we used the Adam optimizer with $betas=(0.9, 0.999)$ and a weight decay of $1e-3$. The loss used was Cross Entropy Loss. Depending upon the dataset, we varied the learning rate and number of epochs by performing an appropriate hyperparamter search and monitoring the training and validation cross entropy losses per epoch. We do not delve into the details for each model.

\begin{table*}[ht]
\centering
\begin{adjustbox}{max width=1\linewidth}
\begin{tabular}{|c|c|c|c|c|} 
\hline
Dataset & Test Accuracy (Unencrypted) & Test Accuracy (Encrypted)  &  Total Inference time (in hrs) & Poly mod degree, Coeff. modulus \\ 
 \hline
  \hline
 MNIST & \textbf{97.14\%}& \textbf{98.90\%} & 1.23 & 8192, (31, 26)\\
  \hline
 Kuzushiji MNIST & \textbf{90.65\%} & \textbf{90.71\%} & 1.31 & 8192,  (31, 26)\\ 
  \hline
 Fashion MNIST & \textbf{87.40\%} & \textbf{87.39\%} & 1.32 & 8192,  (31, 26)\\ 
%
  \hline
\end{tabular}
    \end{adjustbox}
\caption{\label{tab:corruption}{Classification accuracies on test set  (5k samples)  for all datasets  on \textbf{Architecture 1}: a simple 1-convolutional layer CNN, which operates on grayscale images}. The coefficient modulus is a tuple of the number of bits for the fractional+integral and the fractional part respectively.}
\end{table*}

\begin{table*}[ht]
\centering
\begin{adjustbox}{max width=1\linewidth}
\begin{tabular}{|c|c|c|c|c|} 
\hline
Dataset & Test Accuracy (Unencrypted) & Test Accuracy (Encrypted)  &  Total Inference time (in hrs) & Poly mod degree, Coeff. modulus \\ 
 \hline
  \hline
 MNIST & \textbf{94.72\%}& \textbf{94.78\%} & 4.27 & 16384, ((45,30)\\
  \hline
 Kuzushiji MNIST & \textbf{88.46\%} & \textbf{87.66\%} & 4.01 & 16384, (45,30)\\ 
  \hline
 Fashion MNIST & \textbf{84.42\%}& \textbf{83.90\%} & 4.16 & 16384, (31, 26)\\ 
  \hline
\end{tabular}
    \end{adjustbox}
\caption{\label{tab:corruption}{Classification accuracies on test set (5k samples) for all datasets  on \textbf{Architecture 2}: a CNN with 2-convolutional layers, average pooling, and 2 FC layers, which operates on grayscale images. }}
\end{table*}

\begin{table}[h]
\begin{center}
\begin{tabular}{|c||c|}
\hline
Dataset & \textbf{CIFAR-10}\\ \hline
Training Accuracy (Unencrypted) & \textbf{83.9\%} \\ \hline
Test Accuracy (Unencrypted) & \textbf{56.12\%} \\ \hline
Test Accuracy (Encrypted) & \textbf{54.1\%} \\ \hline
Poly mod degree, Coeff. modulus & 16384, (40, 29)\\ \hline
Inference Time (seconds/sample) & \textbf{7.57 s/it}\\ \hline
\end{tabular}
\end{center}
\caption{Classification accuracies for \textbf{1000} CIFAR-10 colored images (3 input channels) on \textbf{Architecture 3}.}
\label{table:specs}
\end{table}

\section{Results and Discussion}

In this section, we discuss the results as presented below in Tables 2 and 3. We present results for classification accuracies and inference times for running test time inference on encrypted neural nets, and compare these to baselines accuracies of the unencrypted counterparts. In the first case for Architecture 1, there is clearly little to no drop off for test time accuracy on encrypted neural networks. In fact, in some cases there is even an improvement. The reason for this could be attributed to the CKKS encryption scheme. Because it is based on approximate arithmetic, the output of the encrypted net is slightly different to the trained model, which adds some natural noise to the predictions. This could reduce overfitting and thus improve test time predictions. In the case of Architecture 2, we see that test time accuracies are lower for KMNIST and Fashion MNIST by approximately 1 percent. The reason for this is that we are performing 1 more convolution and 1 average pooling operation and thus have three sets of encryption and decryption, as opposed to only one. Each time we decrypt the data, the output is slightly different and this may have compounded in the case of Architecture 2. Moreover, we see that the inference times are approximately three times longer in the second case. This is in line with our expectations as adding the extra convolution and pooling operation should scale the timing linearly in the number of encrypted operations taking place. 

\subsection{Context Parameters}
In this section we analyze the impact of the the encryption parameters. In particular, the inference time and the accuracies are affected by the context parameters used in the encryption/decryption. In addition, they also affect the security guarantees provided by the encryption. We analyze these hyperparameters and tradeoffs and try to identify the optimal choice of context. We would also like to point out that the cryptographic nets are run and analyzed on test data and hence the test accuracy is not the classical test accuracy on standard problems and rather serves like training data. In addition, trends in inference time generalize well to a different dataset on similar computer architecture since it is merely dependent on computer architecture. We ran these experiments on the \textit{FashionMNIST} dataset. \\

\begin{figure}
 \begin{center}
 \includegraphics[scale=0.25]{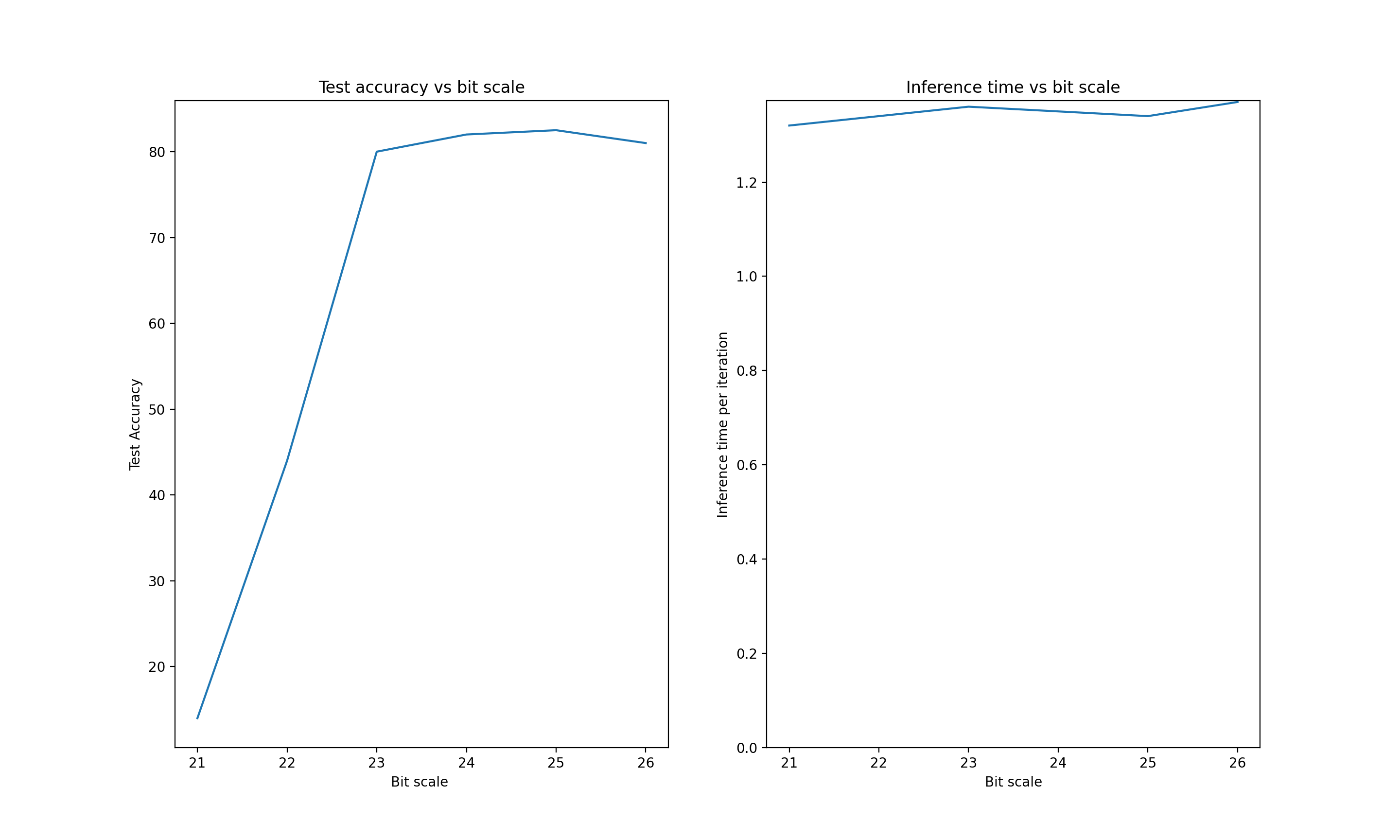}
    \caption{Accuracy and inference time with changing bit scale accuracy for polynomial bit modulus=8192}
     \label{fig:bit_8192}
 \end{center}
 \end{figure}

\begin{figure}
 \begin{center}
 \includegraphics[scale=0.25]{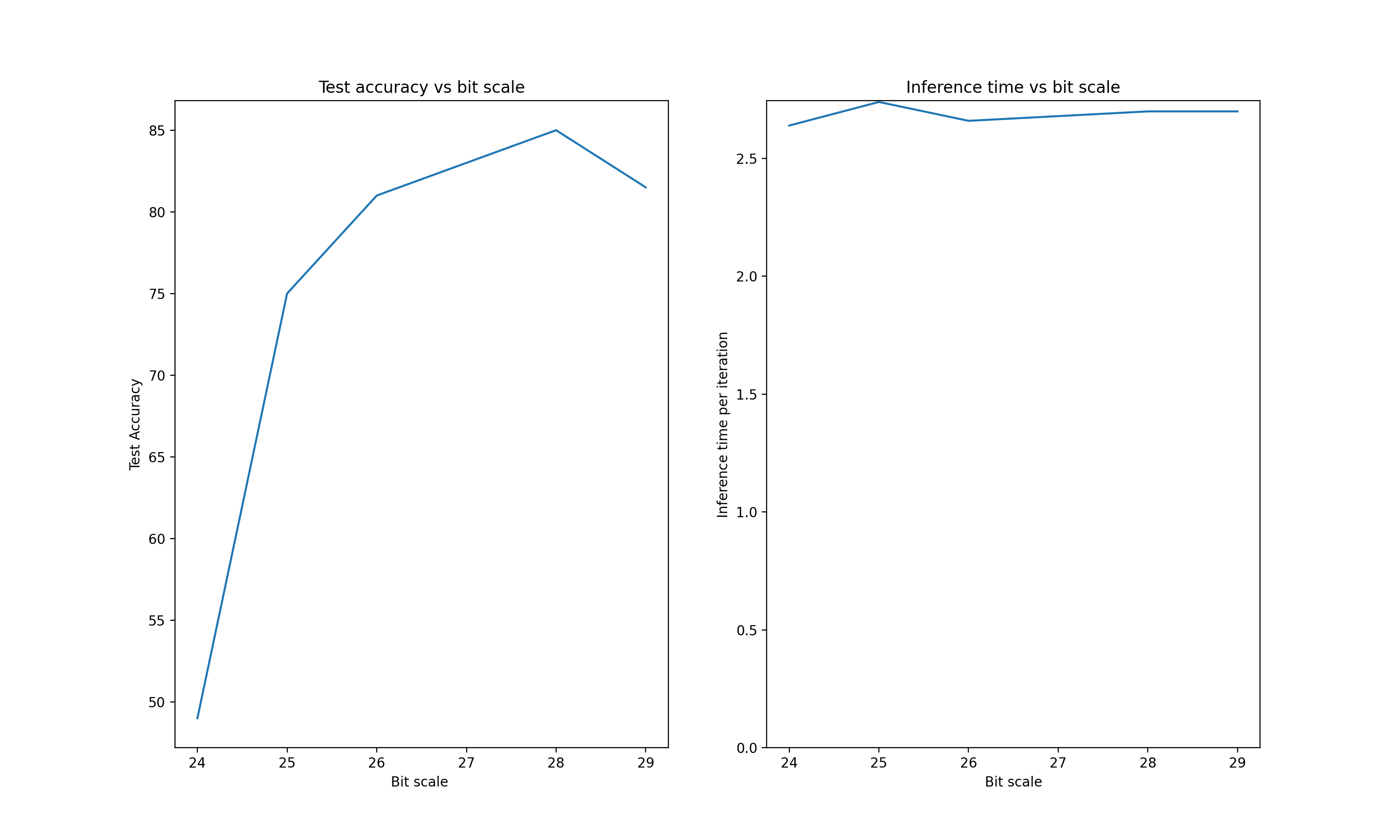}
     \caption{Accuracy and inference time with changing bit scale accuracy for polynomial bit modulus=16384}
     \label{fig:bit_16384}
 \end{center}
 \end{figure}

\textbf{Bit scale:}  We first observe that the reduction in accuracy on encrypted data could be a result of loss of precision during encryption and decryption (we are encrypting and decrypting floating point numbers). The scale controls the precision of the fractional part, since it's the value that plaintexts are multiplied with before being encoded into a polynomial of integer coefficients. We therefore expect that increased bit scale should track the performance of the unencrypted net . However, it is also possible that due to minor errors in encryption, the encrypted net may actually end up classifying an object correctly that it was unable to do so earlier. Hence we analyze the accuracy in Figures \ref{fig:bit_8192} and \ref{fig:bit_16384} and observe that above a certain bit scale, the accuracy remains almost constant and very low below that threshold. We also have 5 bits for the integer part in the coefficient modulus, which should be enough for our use case, since output values are only in the range 1 to 10. In addition, we also expected that the inference time to increase with increasing bit scale. However, as we show in Figures \ref{fig:bit_8192} and \ref{fig:bit_16384}, inference time remains almost constant with increasing bit scale. Hence we conclude that using a high bit scale is beneficial for encryption (we use 26 bit bit scale for polynomial bit modulus=8192).\\

\textbf{Polynomial bit modulus}: A higher polynomial bit modulus stands for stronger security guarantees, though it results in increased inference time. Tenseal library imposes certain constraints on the relation between polynomial bit modulus and bit scale (we cannot run more than 26 bits for 8192 polynomial bit modulus to ensure security guarantees). To run bigger bit scales, we double the bit modulus and analyze its impact on accuracy and inference time. We observe that the inference time increases in proportion, while the accuracy remains similar. Thus, this represents a tradeoff between security guarantees and inference time: to ensure strong security guarantees, we may be forced to increase the inference time.
\begin{table}[h]
\begin{center}
\begin{tabular}{|c||c|}
\hline
Polynomial Bit Modulus & Inference time (s/it)\\ \hline
8192 & \textbf{1.31} \\ \hline
16384 & \textbf{2.62} \\ \hline
\end{tabular}
\end{center}
\caption{Inference time doubles when we increase the polynomial bit modulus: This represents a tradeoff between security guarantee and inference time.}
\label{table:specs}
\end{table} 

\section{Conclusion}
The lack of GPU support in Microsoft SEAL based libraries makes the encrypted-CNNs very slow. As noted in Section 3, adding parallel computing and substituting implementations of faster Homomorphic Encryption scheme  in libraries like Microsoft SEAL and TenSEAL can greatly improve performance of these encrypted-CNNs. Another observation is that choosing the encryption parameters is often not very easy and requires some intuition and tuning, much like tuning the other hyperparameters of the neural network.
However, this is an exciting field of research since there is great value in being able to perform Computer Vision tasks directly on cipher-texts to tackle serious privacy concerns. We demonstrated some techniques, implemented few useful layers on top of an open-source library to achieve this and extended these techniques to new and more complex datasets.

{\small
\bibliographystyle{ieee}
\bibliography{ref}

@misc{https://doi.org/10.48550/arxiv.1412.6181,
  doi = {10.48550/ARXIV.1412.6181},
  
  url = {https://arxiv.org/abs/1412.6181},
  
  author = {Xie, Pengtao and Bilenko, Misha and Finley, Tom and Gilad-Bachrach, Ran and Lauter, Kristin and Naehrig, Michael},
  
  title = {Crypto-Nets: Neural Networks over Encrypted Data},
  
  publisher = {arXiv},
  
  year = {2014},
  
  copyright = {arXiv.org perpetual, non-exclusive license}
}

@misc{sealcrypto,
    title = {{M}icrosoft {SEAL} (release 4.0)},
    howpublished = {\url{https://github.com/Microsoft/SEAL}},
    month = mar,
    year = 2022,
    note = {Microsoft Research, Redmond, WA.},
    key = {SEAL}
}

@misc{pysyft,
  doi = {10.48550/ARXIV.1811.04017},
  
  url = {https://arxiv.org/abs/1811.04017},
  
  author = {Ryffel, Theo and Trask, Andrew and Dahl, Morten and Wagner, Bobby and Mancuso, Jason and Rueckert, Daniel and Passerat-Palmbach, Jonathan},
  
  title = {A generic framework for privacy preserving deep learning},
  
  publisher = {arXiv},
  
  year = {2018},
  
  copyright = {arXiv.org perpetual, non-exclusive license}
}

@misc{tenseal2021,
    title={TenSEAL: A Library for Encrypted Tensor Operations Using Homomorphic Encryption}, 
    author={Ayoub Benaissa and Bilal Retiat and Bogdan Cebere and Alaa Eddine Belfedhal},
    year={2021},
    eprint={2104.03152},
    archivePrefix={arXiv},
    primaryClass={cs.CR}
}

@article{Rivest1978,
  author = {Rivest, R L and Adleman, L and Dertouzos, M L},
  }

@inproceedings{gentry,
author = {Gentry, Craig},
title = {Fully Homomorphic Encryption Using Ideal Lattices},
year = {2009},
isbn = {9781605585062},
publisher = {Association for Computing Machinery},
address = {New York, NY, USA},
url = {https://doi.org/10.1145/1536414.1536440},
doi = {10.1145/1536414.1536440},
booktitle = {Proceedings of the Forty-First Annual ACM Symposium on Theory of Computing},
pages = {169–178},
numpages = {10},
location = {Bethesda, MD, USA},
series = {STOC '09}
}

@inproceedings{naehrig,
author = {Naehrig, Michael and Lauter, Kristin and Vaikuntanathan, Vinod},
year = {2011},
month = {10},
pages = {113-124},
title = {Can Homomorphic Encryption be Practical?},
journal = {Proceedings of the 3rd ACM Workshop on Cloud Computing Security Workshop},
doi = {10.1145/2046660.2046682}
}

@inproceedings{liu2015faceattributes,
  title = {Deep Learning Face Attributes in the Wild},
  author = {Liu, Ziwei and Luo, Ping and Wang, Xiaogang and Tang, Xiaoou},
  booktitle = {Proceedings of International Conference on Computer Vision (ICCV)},
  month = {December},
  year = {2015} 
}

@article{fl,
author = {Yang, Qiang and Liu, Yang and Chen, Tianjian and Tong, Yongxin},
title = {Federated Machine Learning: Concept and Applications},
year = {2019},
issue_date = {March 2019},
publisher = {Association for Computing Machinery},
address = {New York, NY, USA},
volume = {10},
number = {2},
issn = {2157-6904},
url = {https://doi.org/10.1145/3298981},
doi = {10.1145/3298981},
journal = {ACM Trans. Intell. Syst. Technol.},
month = {jan},
articleno = {12},
numpages = {19}
}

@ARTICLE{he,  author={Phong, Le Trieu and Aono, Yoshinori and Hayashi, Takuya and Wang, Lihua and Moriai, Shiho},  journal={IEEE Transactions on Information Forensics and Security},   title={Privacy-Preserving Deep Learning via Additively Homomorphic Encryption},   year={2018},  volume={13},  number={5},  pages={1333-1345},  doi={10.1109/TIFS.2017.2787987}}

@inproceedings{rw,
author = {Disabato, Simone and Falcetta, Alessandro and Mongelluzzo, Alessio and Roveri, Manuel},
year = {2020},
month = {07},
pages = {1-8},
title = {A Privacy-Preserving Distributed Architecture for Deep-Learning-as-a-Service},
doi = {10.1109/IJCNN48605.2020.9207619}
}

@article{DBLP:journals/corr/LiEKNKT16,
  author    = {Yunzhu Li and
               Andre Esteva and
               Brett Kuprel and
               Roberto A. Novoa and
               Justin Ko and
               Sebastian Thrun},
  title     = {Skin Cancer Detection and Tracking using Data Synthesis and Deep Learning},
  journal   = {CoRR},
  volume    = {abs/1612.01074},
  year      = {2016},
  url       = {http://arxiv.org/abs/1612.01074},
  eprinttype = {arXiv},
  eprint    = {1612.01074},
  bibsource = {dblp computer science bibliography, https://dblp.org}
}

@InProceedings{Wang_2020_CVPR,
author = {Wang, Dong and Zhang, Yuan and Zhang, Kexin and Wang, Liwei},
title = {FocalMix: Semi-Supervised Learning for 3D Medical Image Detection},
booktitle = {Proceedings of the IEEE/CVF Conference on Computer Vision and Pattern Recognition (CVPR)},
month = {June},
year = {2020}
}

@article{brain_tumor,
  author    = {Amin, J. and Sharif, M. and Haldorai, A.},
  title     = {Brain tumor detection and classification using machine learning: a comprehensive survey},
  journal   = {Complex Intell. Syst.},
  year      = {2021},
  url       = {https://doi.org/10.1007/s40747-021-00563-y}
}

@article{fast_HME,
  title={TFHE: Fast Fully Homomorphic Encryption Over the Torus},
  journal={Journal of Cryptology},
  publisher={Springer},
  volume={},
  doi={10.1007/s00145-019-09319-x},
  author={Ilaria Chillotti and Nicolas Gama and Mariya Georgieva and Malika Izabachène},
  year=2019
}

@misc{fast_hme_f1,
  doi = {10.48550/ARXIV.2109.05371},
  
  url = {https://arxiv.org/abs/2109.05371},
  
  author = {Feldmann, Axel and Samardzic, Nikola and Krastev, Aleksandar and Devadas, Srini and Dreslinski, Ron and Eldefrawy, Karim and Genise, Nicholas and Peikert, Chris and Sanchez, Daniel},
  
  title = {F1: A Fast and Programmable Accelerator for Fully Homomorphic Encryption (Extended Version)},
  
  publisher = {arXiv},
  
  year = {2021},
  
  copyright = {Creative Commons Attribution Non Commercial Share Alike 4.0 International}
}

@inproceedings{crypto-2013-24633,
  title={Homomorphic Encryption from Learning with Errors: Conceptually-Simpler, Asymptotically-Faster, Attribute-Based},
  booktitle={CRYPTO},
  publisher={Springer},
  pages={75-92},
  url={https://www.iacr.org/archive/crypto2013/80420297/80420297.pdf},
  doi={10.1007/978-3-642-40041-4_5},
  author={Craig Gentry and Amit Sahai and Brent Waters},
  year=2013
}

@article{LpezAlt2012OntheflyMC,
  title={On-the-fly multiparty computation on the cloud via multikey fully homomorphic encryption},
  author={Adriana L{\'o}pez-Alt and Eran Tromer and Vinod Vaikuntanathan},
  journal={IACR Cryptol. ePrint Arch.},
  year={2012},
  volume={2013},
  pages={94}
}

@inproceedings{torchvision,
author = {Marcel, S\'{e}bastien and Rodriguez, Yann},
title = {Torchvision the Machine-Vision Package of Torch},
year = {2010},
isbn = {9781605589336},
publisher = {Association for Computing Machinery},
address = {New York, NY, USA},
url = {https://doi.org/10.1145/1873951.1874254},
doi = {10.1145/1873951.1874254},
booktitle = {Proceedings of the 18th ACM International Conference on Multimedia},
pages = {1485–1488},
numpages = {4},
location = {Firenze, Italy},
series = {MM '10}
}
}

\end{document}